\documentclass[twocolumn,aps,prd,preprintnumbers,showpacs,superscriptaddress,nofootinbib,amsmath,amssymb,floats,floatfix,showkeys,notitlepage,longbibliography]{revtex4-2}
\usepackage{subcaption}
\usepackage{graphicx}
\usepackage{xcolor}
\usepackage{hyperref}
\hypersetup{colorlinks=true,linkcolor=blue,urlcolor=blue,citecolor=blue}
\usepackage{orcidlink}
\usepackage{lipsum}
\usepackage{etoolbox}

\begin{document}

\setcounter{MaxMatrixCols}{10}

\title{Regular Magnetically Charged Black Holes from Nonlinear Electrodynamics: Thermodynamics, Light Deflection, and Orbital Dynamics}

\author{Ekrem Aydiner \orcidlink{0000-0002-0385-9916}}
\email{aydiner@princeton.edu}
\affiliation{Department of Physics, Princeton University, Princeton, New Jersey 08544, USA}
\affiliation{Department of Physics, \.{I}stanbul University, \.{I}stanbul 34134, T\"{u}rkiye}

\author{Erdem Sucu
\orcidlink{0009-0000-3619-1492}}
\email{erdemsc07@gmail.com}
\affiliation{Physics Department, Eastern Mediterranean
University, Famagusta, 99628 North Cyprus, via Mersin 10, T\"{u}rkiye}

\author{\.{I}zzet Sakall{\i} \orcidlink{0000-0001-7827-9476}}
\email{izzet.sakalli@emu.edu.tr}
\affiliation{Physics Department, Eastern Mediterranean
University, Famagusta, 99628 North Cyprus, via Mersin 10, T\"{u}rkiye}

\begin{abstract}
We investigate the thermodynamic properties, light deflection, and orbital dynamics of regular magnetically charged black holes (NRCBHs) arising from nonlinear electrodynamics (NED) coupled to general relativity. The metric function $f(r)$ ensures complete regularity at the origin while maintaining asymptotic flatness, with the extremal magnetic charge limit reaching $q_{\text{ext}} \approx 2.54M$, significantly exceeding the Reissner-Nordstr\"{o}m value. Using the quantum tunneling framework, we derive the Hawking temperature and incorporate generalized uncertainty principle (GUP) corrections, showing $T_{\text{GUP}} = (f'(r_h)/4\pi)\sqrt{1-2\beta m_p^2}$. The weak deflection of light is analyzed through the Gauss-Bonnet theorem (GBT), revealing charge-dependent behavior where large $q$ values lead to negative deflection angles due to electromagnetic repulsion. Plasma effects further modify the deflection through the refractive index $n(r) = \sqrt{1 - \omega_p^2(r)f(r)/\omega_0^2}$. Keplerian motion analysis demonstrates that the angular velocity $\Omega(r)$ exhibits charge-sensitive maxima related to quasi-periodic oscillations (QPOs) in accretion disks. Finally, we examine Joule-Thomson expansion (JTE) properties, finding that the coefficient $\mu_J$ indicates cooling behavior for higher charges and larger event horizons. Our results provide comprehensive insights into the observational signatures of NRCBHs, with implications for gravitational lensing, X-ray astronomy, and tests of nonlinear electromagnetic theories in strong gravitational fields.
\end{abstract}

\date{June 21, 2025}


\maketitle
\tableofcontents

\section{Introduction} \label{isec1}

Black holes (BHs) represent one of the most fascinating and enigmatic predictions of Einstein's general relativity, serving as natural laboratories for testing our understanding of fundamental physics in extreme gravitational environments \cite{berti2013astrophysical,yagi2016black,bambi2018astrophysical}. Among the diverse family of BH solutions, regular BHs (RBHs) have emerged as particularly compelling objects that address the longstanding issue of spacetime singularities while maintaining the essential characteristics of classical BHs \cite{colleaux2018nonpolynomial,sebastiani2022some,kar2024novel}. Unlike their singular counterparts, RBHs possess a smooth, geodesically complete spacetime structure that eliminates the problematic infinities at the center, typically replacing the singular core with a de Sitter-like region or other regular geometries \cite{lemos2011regular,bouhmadi2021regular}. This regularity is often achieved through modifications of Einstein's field equations or by coupling gravity to nonlinear matter fields that provide the necessary energy-momentum sources to support such exotic geometries.

The concept of magnetically charged BHs from NED has gained significant attention in recent years, as these solutions offer a natural pathway to constructing regular spacetimes while maintaining electromagnetic duality and preserving the fundamental principles of general relativity \cite{allahyari2020magnetically,wen2023observational,stuchlik2019generic}. Nonlinear electromagnetic theories, which extend Maxwell's linear electrodynamics, have deep roots in both classical and quantum field theory, arising from effective descriptions of quantum chromodynamics, string theory corrections, and various unified field theories \cite{fradkin1985non}. When coupled to Einstein's gravity, NED models can generate regular BH solutions where the nonlinear electromagnetic stress-energy tensor provides the exotic matter required to smooth out the central singularity \cite{ayon1999new,balart2014regular,bronnikov2018nonlinear}. These magnetically charged configurations are particularly interesting because they preserve electromagnetic duality while exhibiting enhanced stability properties compared to their electrically charged counterparts, especially in astrophysical environments where charge neutralization typically occurs rapidly \cite{breton2005stability,gonzalez2009thermodynamics}.

BH thermodynamics represents one of the most profound connections between gravity, quantum mechanics, and statistical physics, establishing fundamental relationships between geometric properties of BH spacetimes and thermodynamic quantities \cite{gecim2017gup,wald1999gravitation,yasir2023thermal,gecim2020quantum,gursel2025thermodynamics}. The pioneering work of Bekenstein and Hawking revealed that BHs possess intrinsic temperature and entropy, with the latter being proportional to the area of the event horizon rather than the volume, suggesting a holographic nature of gravitational degrees of freedom \cite{bekenstein2008bekenstein}. For charged BHs, the thermodynamic description becomes richer, incorporating electric or magnetic charge as an additional extensive variable analogous to chemical potential in ordinary thermodynamic systems \cite{kubizvnak2012p,gibbons1977action}. The study of thermodynamic properties in regular charged BHs is particularly fascinating because the absence of singularities allows for a more complete understanding of the microscopic degrees of freedom and their statistical mechanical interpretation \cite{myung2009thermodynamics}. Recent developments have also focused on the role of quantum gravitational effects, such as those arising from the GUP, which introduce corrections to the standard Hawking temperature and provide insights into the quantum nature of gravity near the Planck scale \cite{adler2001generalized,nozari2006comparison,gecim2018gup,pedram2011effects}.

The JTE in BH physics has emerged as a powerful tool for investigating phase transitions and critical phenomena in gravitational thermodynamics \cite{okcu2017joule,okcu2018joule}. This process, which involves isenthalpic expansion or compression, allows for the identification of cooling and heating regimes characterized by the sign of the Joule-Thomson coefficient $\mu_J$ \cite{lan2018joule}. In the context of BH thermodynamics within extended phase space, where the cosmological constant is interpreted as thermodynamic pressure and its conjugate quantity as thermodynamic volume, JTE provides crucial insights into the stability and phase structure of BH solutions. In charged black holes, the JTE analysis gains depth from the interaction between electromagnetic and gravitational influences on thermodynamic potentials, frequently uncovering intricate phase diagrams with numerous critical points and unique phase transitions \cite{adimurthi2018nonlinear}.

Light deflection by massive objects represents one of the most celebrated predictions and confirmations of general relativity, with observations of stellar light bending around the Sun during solar eclipses providing early experimental validation of Einstein's theory. The study of photon trajectories in BH spacetimes has evolved into a sophisticated field that combines analytical techniques from differential geometry with computational methods to understand gravitational lensing phenomena \cite{perlick2004gravitational,sucu2025probing,virbhadra2002gravitational,bozza2002gravitational}. The GBT has proven particularly valuable for calculating weak deflection angles in various BH geometries, providing exact expressions for the bending of light in terms of the curvature properties of the optical metric \cite{gibbons2008applications,sucu2024effect,jusufi2018deflection,kacsikcci2019gravitational}. In charged black holes, the electromagnetic field complicates photon dynamics, balancing gravitational attraction with electromagnetic repulsion, potentially yielding negative deflection angles in specific scenarios \cite{heydari2022null,jusufi2016gravitational,sucu2025exploring}. The incorporation of plasma effects further enriches this picture, as astrophysical environments typically contain ionized matter that modifies the effective refractive index and introduces frequency-dependent corrections to light propagation \cite{Perlick:2015vta,tsupko2017analytical}.

Orbital dynamics around BHs provides fundamental insights into the behavior of test particles and extended objects in strong gravitational fields, with applications ranging from accretion disk physics to gravitational wave astronomy \cite{bardeen1972rotating,shapiro2024black,sucu2025dynamics}. The study of circular orbits, in particular, reveals characteristic frequencies that are directly observable in astrophysical systems through QPOs in X-ray emissions from accreting BHs \cite{stella1997lense,remillard2006x}. For charged BHs, the presence of electromagnetic fields modifies the effective potential experienced by charged test particles, leading to rich orbital structures that depend sensitively on the charge-to-mass ratios of both the BH and the orbiting particle \cite{pugliese2011motion,konoplya2006particle}. The Keplerian angular velocity and its charge-dependent modifications provide crucial information for distinguishing between different BH models and testing theories of modified gravity in strong-field regimes \cite{sucu2025nonlinear}.

Our primary motivation for this comprehensive study stems from the recognition that NRCBHs represent a unique class of solutions that simultaneously address several fundamental questions in theoretical physics and astrophysics. First, these solutions provide a natural resolution to the singularity problem while maintaining physically reasonable energy conditions and electromagnetic duality. Second, they offer a rich laboratory for exploring quantum gravitational effects through modified thermodynamic properties and their corrections under frameworks such as GUP. Third, their distinctive observational signatures in light deflection and orbital dynamics make them potentially distinguishable from other BH models, providing avenues for experimental tests of NED theories and regular BH concepts.The specific aims of our investigation are multifold: (i) to provide a comprehensive analysis of the thermodynamic properties of NRCBHs, including Hawking temperature calculations using quantum tunneling methods and their modifications under GUP corrections; (ii) to examine the weak deflection of light using the GBT approach, incorporating both vacuum and plasma-mediated environments to understand the full range of lensing phenomena; (iii) to analyze the orbital dynamics of neutral test particles, focusing on the characteristic frequencies relevant for QPO observations and their charge-dependent modifications; (iv) to investigate the Joule-Thomson expansion properties and their implications for thermodynamic stability and phase transitions; and (v) to synthesize these results to provide a unified picture of NRCBH observational signatures that could guide future experimental and observational efforts.

The paper is organized as follows: Section~\ref{isec2} reviews the regular charged BH geometry from NED, establishing the mathematical framework and discussing horizon structure, regularity conditions, and energy constraints. Section~\ref{isec3} analyzes weak light deflection in NRCBH spacetimes, applying the GBT method and incorporating plasma effects to derive comprehensive deflection angle formulas. Section~\ref{isec4} investigates Keplerian motion and orbital dynamics, focusing on characteristic frequencies and their relevance to QPO observations. Section~\ref{isec5} examines Joule-Thomson expansion properties, deriving the expansion coefficient and analyzing cooling/heating behavior. Finally, Section~\ref{isec6} presents our conclusions and discusses our findings for BH physics and observational astronomy.

\section{Review of NRCBH Spacetime} \label{isec2}

We consider a static and spherically symmetric BH solution that remains regular throughout the entire spacetime and originates from the coupling of general relativity with a NED model. The line element is taken in the usual form
\begin{equation}
ds^2 = -f(r), dt^2 + \frac{dr^2}{f(r)} + r^2(d\theta^2 + \sin^2\theta, d\phi^2),
\end{equation}
where the metric function $f(r)$ is defined by \cite{balart2025new}
\begin{equation}
f(r) = 1 - \frac{2M}{r} \left( 1 - \frac{M q^2}{M q^2 + 8r^3} + \frac{q^2 r^3}{2M(q^2 + r^2)^2} \right). \label{ismetric}
\end{equation}
Here, $M$ is the gravitational mass and $q$ denotes the magnetic charge. The construction of $f(r)$ ensures regularity at the origin while recovering the correct asymptotic behavior as $r \to \infty$.
\begin{table}[ht]
\renewcommand{\arraystretch}{1.3}
\setlength{\tabcolsep}{10pt}
\begin{center}
\begin{tabular}{|c|c|}
\hline
\textbf{Magnetic Charge ($q$)} & \textbf{Horizon(s)} \\
\hline
0.0 & $[2.0]$ \\
\hline
0.5 & $[0.12523832,\ 2.0998941]$ \\
\hline
0.9 & $[0.23132603,\ 2.2499199]$ \\
\hline
1.0 & $[0.25880556,\ 2.2874627]$ \\
\hline
1.5 & $[0.40309085,\ 2.4482138]$ \\
\hline
1.8 & $[0.49629942,\ 2.5134738]$ \\
\hline
2.0 & $[0.56190919,\ 2.5410464]$ \\
\hline
2.2 & $[0.63086070,\ 2.5545455]$ \\
\hline
2.4 & $[0.70381773,\ 2.5528097]$ \\
\hline
2.5 & $[0.74207114,\ 2.5458337]$ \\
\hline
2.6 & $[0.78169629,\ 2.5345501]$ \\
\hline
2.8 & $[0.86582217,\ 2.4982343]$ \\
\hline
3.0 & $[0.95824262,\ 2.4418624]$ \\
\hline
3.1 & $[1.0085561,\ 2.4052770]$ \\
\hline
3.5 & $[1.2581076,\ 2.1847352]$ \\
\hline
3.6 & $[1.3444531,\ 2.1001215]$ \\
\hline
3.7 & $[1.4572191,\ 1.9873840]$ \\
\hline
3.79 & $[1.6519315,\ 1.7914870]$ \\
\hline
3.8 & $[\mathrm{None}]$ \\
\hline
4.0 & $[\mathrm{None}]$ \\
\hline
\end{tabular}
\caption{Horizon radii $r_h$ obtained from the metric function \eqref{ismetric} of NRCBH with mass $M=1$ for various magnetic charge values $q$. The presence of two distinct roots corresponds to non-extremal BHs, while a single degenerate root ($q=0$) indicates an extremal BH: the Schwarzschild BH. The absence of real positive roots implies that the solution no longer represents a BH, which has a naked singularity at the origin.}
\label{istable}
\end{center}
\end{table}

Table~\ref{istable} reveals the horizon structure evolution of the NRCBH as a function of magnetic charge $q$ for unit mass $M = 1$. For $q = 0$, the solution reduces to the Schwarzschild case with a single horizon at $r_h = 2$. As $q$ increases from $0.5$ to $3.79$, the geometry exhibits two distinct horizons corresponding to non-extremal configurations, with the outer horizon reaching a maximum radius of $r_h \approx 2.55$ at $q = 2.2$ before decreasing toward the extremal limit. The critical transition occurs at $q_{\text{ext}} \approx 3.79$, significantly larger than the Reissner-Nordström extremal charge $q = 1$ (for $M = 1$), demonstrating the stabilizing effect of nonlinear electrodynamics on the horizon structure. Beyond this threshold ($q \geq 3.8$), no real horizons exist, resulting in naked singularities that violate cosmic censorship, thus establishing the physical bounds for viable NRCBH solutions.
\begin{figure*}
    \centering
    \setlength{\tabcolsep}{0pt} 
    \begin{minipage}{0.28\textwidth}
        \centering
        \includegraphics[width=\textwidth]{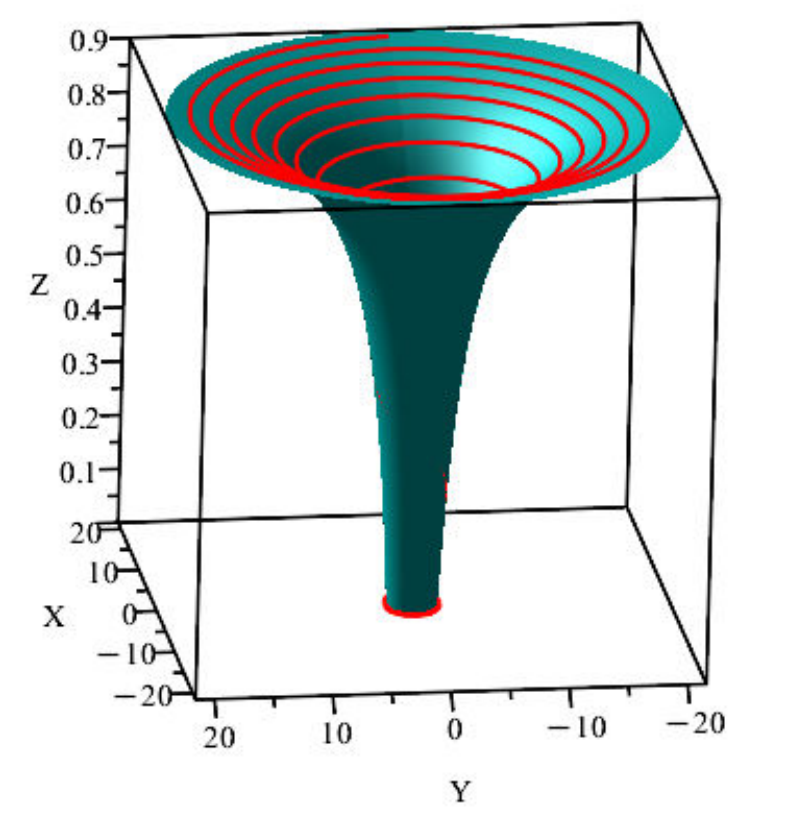}
        \subcaption{[$q=0$, $r_h=2$] \newline NRCBH without magnetic charge: Schwarzchild BH case.}
        \label{fig:is1}
    \end{minipage}
    \begin{minipage}{0.28\textwidth}
        \centering
        \includegraphics[width=\textwidth]{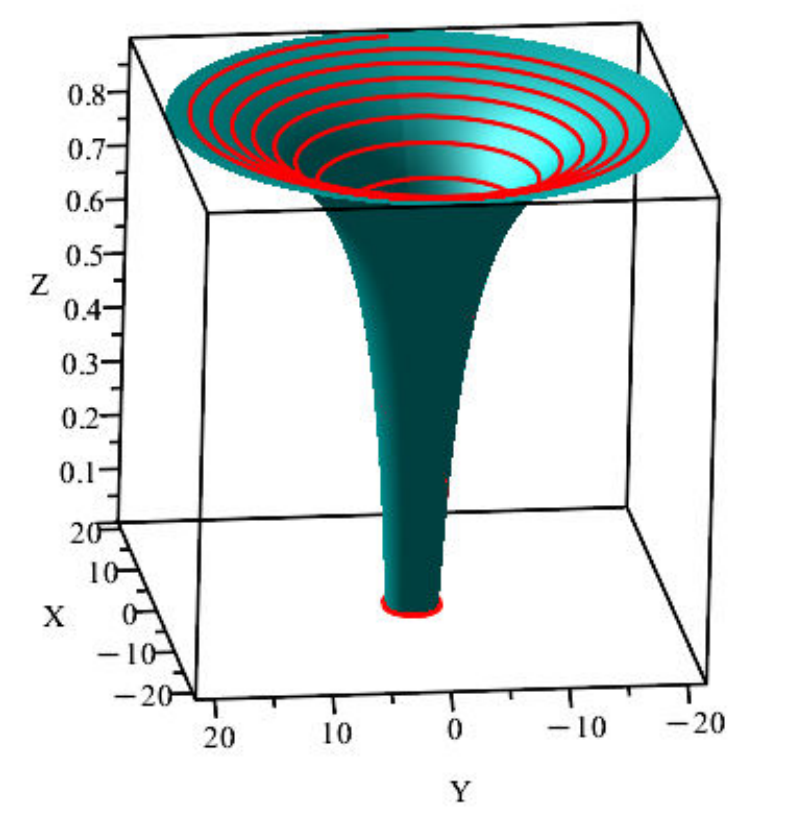}
        \subcaption{[$q=0.5$, $r_h=2.0998941$] \newline NRCBH with magnetic charge.}
        \label{fig:is2}
    \end{minipage}
    \begin{minipage}{0.28\textwidth}
        \centering
        \includegraphics[width=\textwidth]{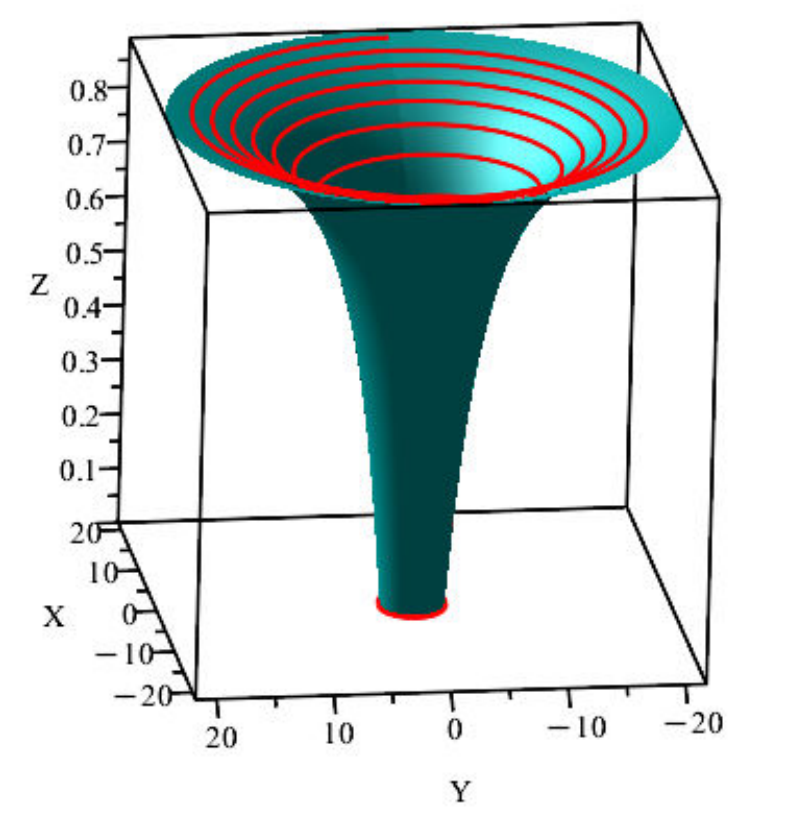}
        \subcaption{[$q=2$, $r_h=2.5410464$] \newline RCBH with magnetic charge.}
        \label{fig:is3}
    \end{minipage}

    \vspace{0.5em} 

    \begin{minipage}{0.28\textwidth}
        \centering
        \includegraphics[width=\textwidth]{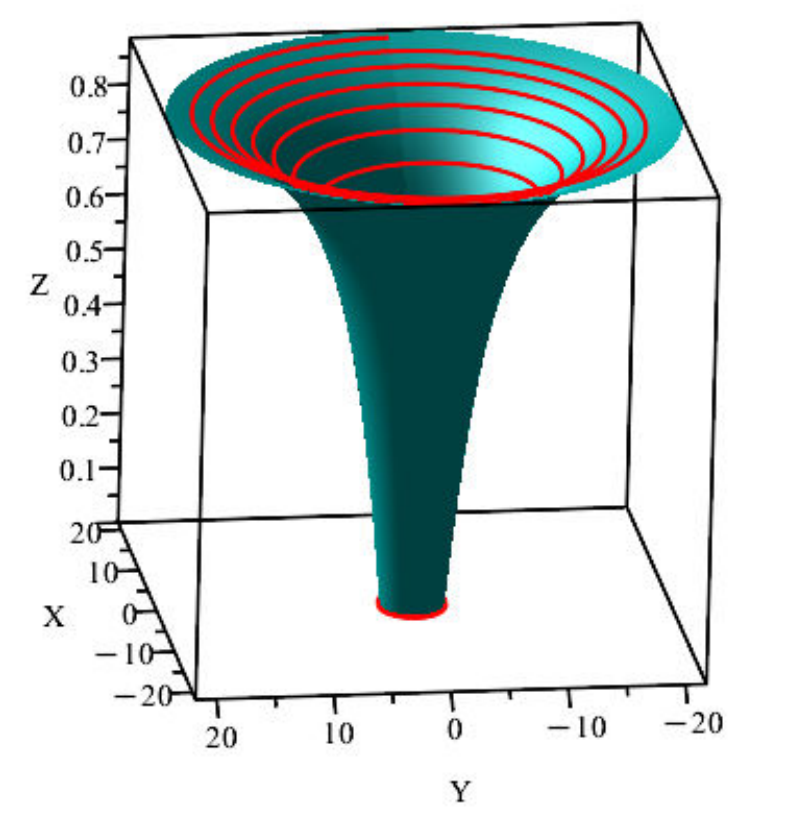}
        \subcaption{[$q=2.5$, $r_h=2.5458337$] \newline NRCBH with magnetic charge.}
        \label{fig:is4}
    \end{minipage}
    \begin{minipage}{0.28\textwidth}
        \centering
        \includegraphics[width=\textwidth]{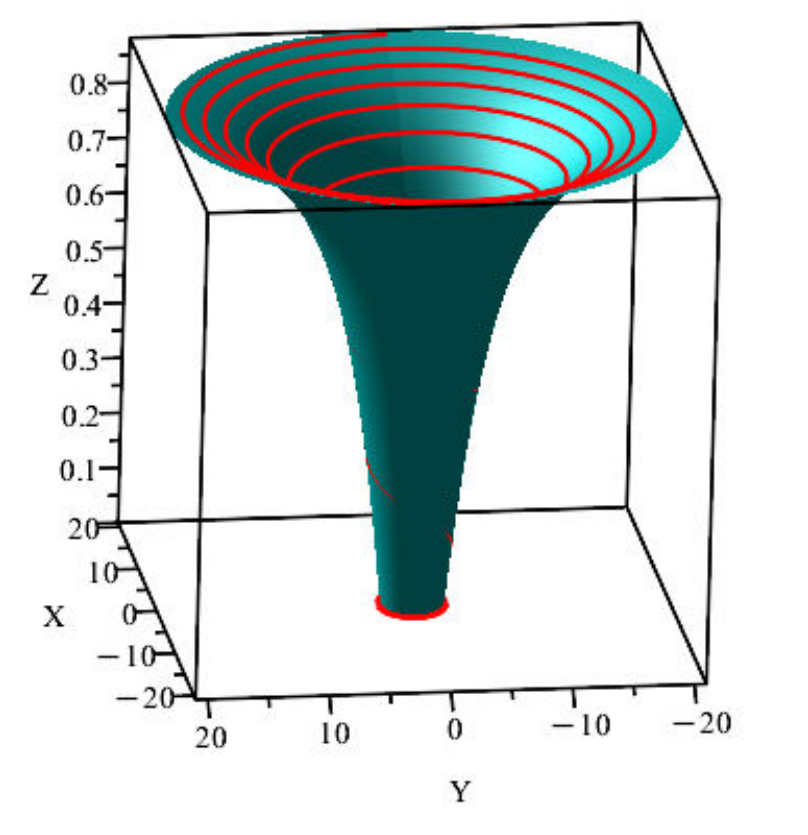}
        \subcaption{[$q=2.8$, $r_h=2.4982343$] \newline NRCBH with magnetic charge.}
        \label{fig:is5}
    \end{minipage}
    \begin{minipage}{0.28\textwidth}
        \centering
        \includegraphics[width=\textwidth]{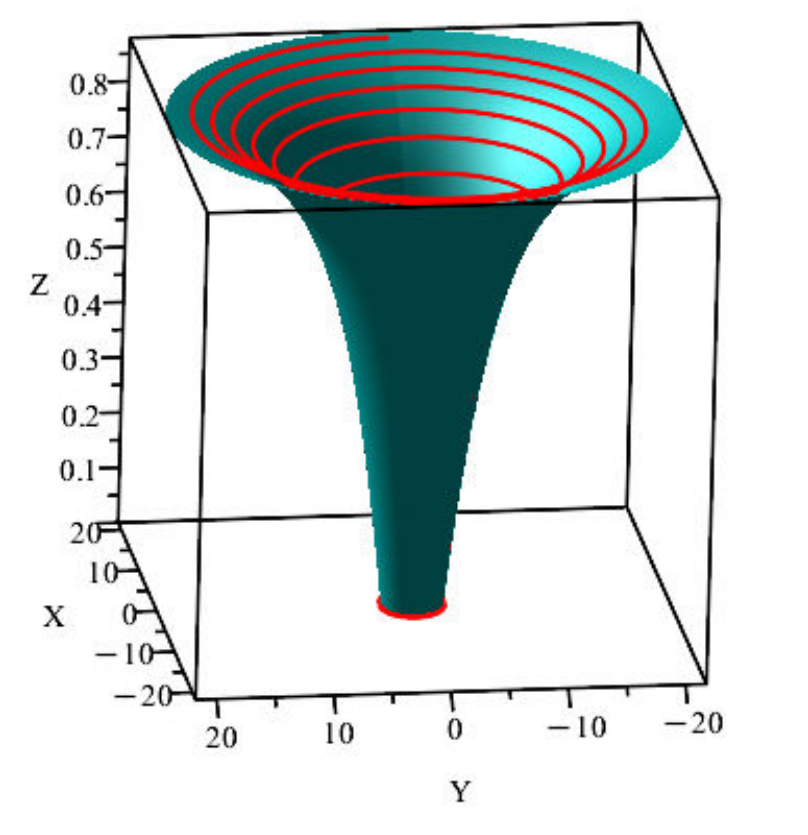}
        \subcaption{[$q=3$, $r_h=2.4418624$] \newline NRCBH with magnetic charge.}
        \label{fig:is6}
    \end{minipage}
    \vspace{0.5em}
    \begin{minipage}{0.28\textwidth}
       \centering
        \includegraphics[width=\textwidth]{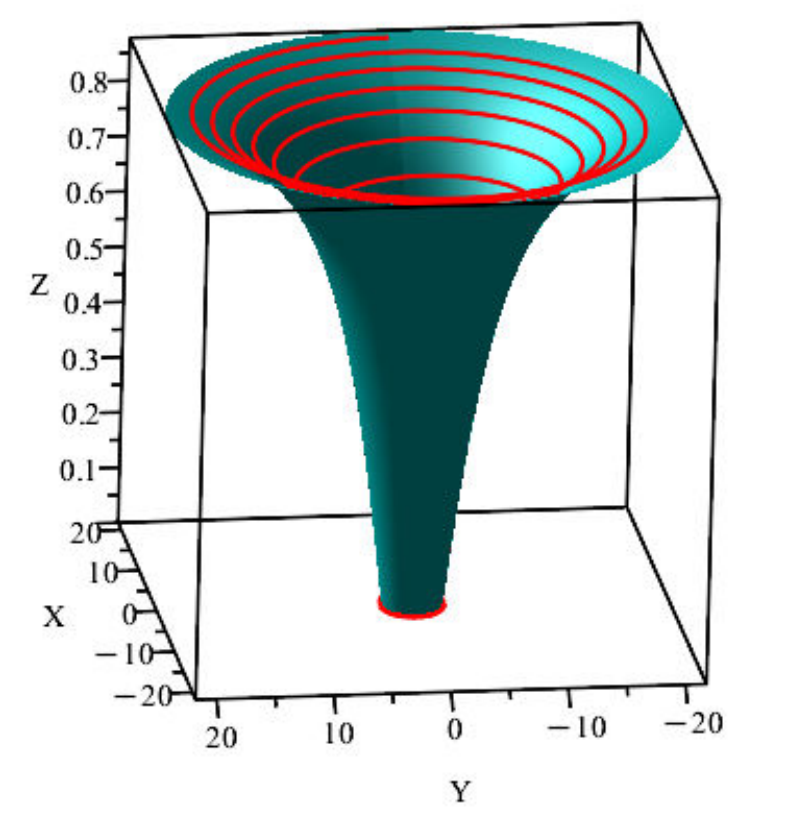}
        \subcaption{[$q=3.1$, $r_h=2.4052770$] \newline NRCBH with magnetic charge.}
        \label{fig:is7}
    \end{minipage}
    \begin{minipage}{0.28\textwidth}
       \centering
        \includegraphics[width=\textwidth]{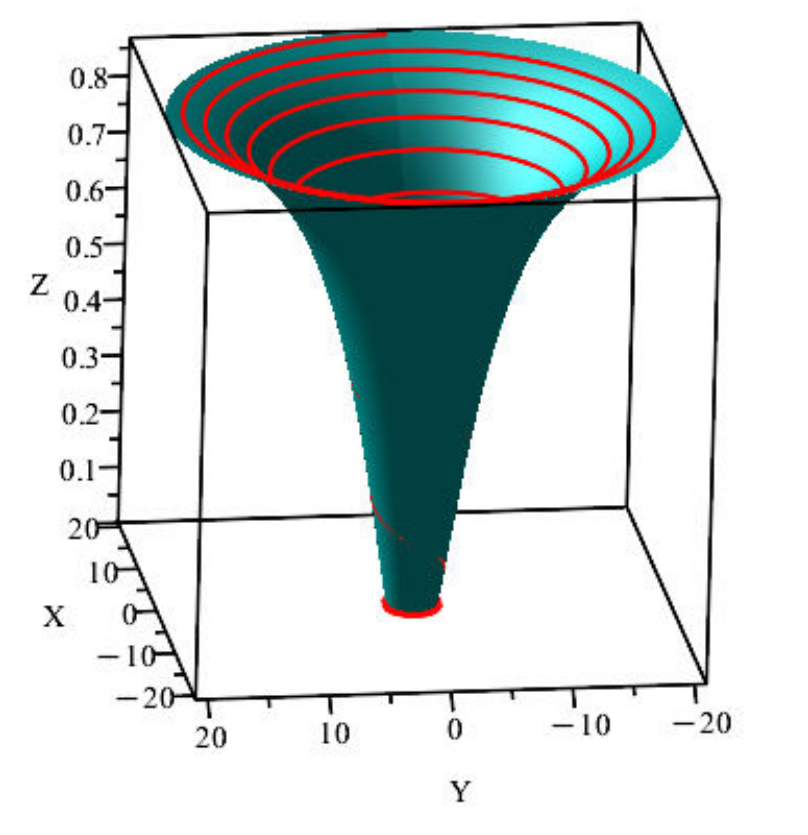}
        \subcaption{[$q=3.7$, $r_h=1.9873840$] \newline RCBH with magnetic charge.}
        \label{fig:is8}
    \end{minipage}
    \begin{minipage}{0.28\textwidth}
       \centering
        \includegraphics[width=\textwidth]{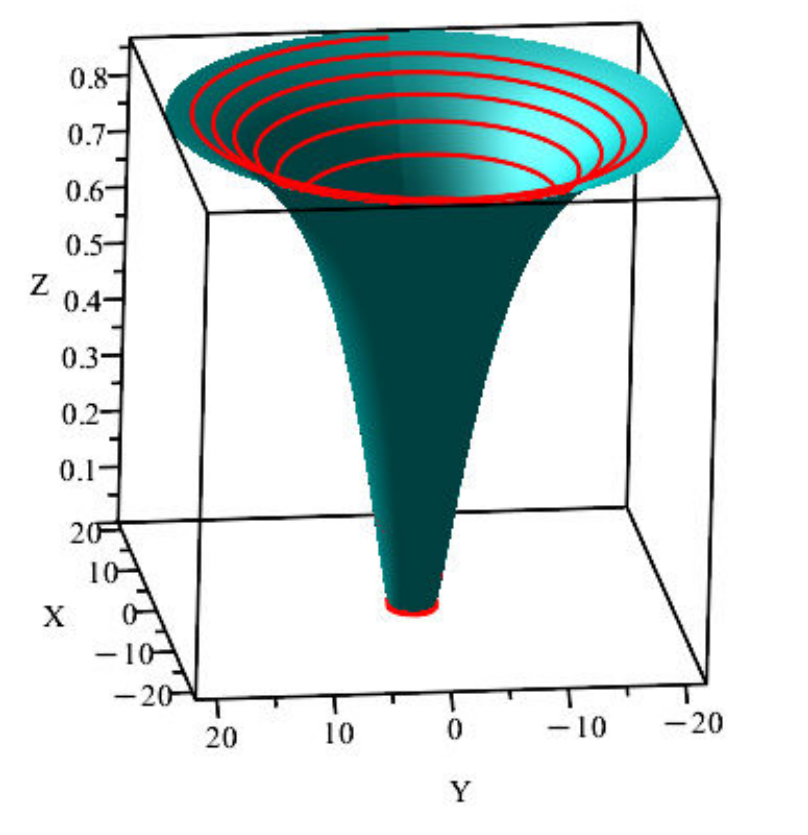}
        \subcaption{[$q=3.79$, $r_h=1.7914870$] \newline NRCBH with magnetic charge.}
        \label{fig:is9}
    \end{minipage}
    \caption{Embedding diagrams of the NRCBH for various values of the Hernquist-DM parameters. The BH mass is set to $M=1$.}
    \label{figisfull}
\end{figure*}
\begin{figure}[h]
    \centering
    \includegraphics[width=0.48\textwidth]{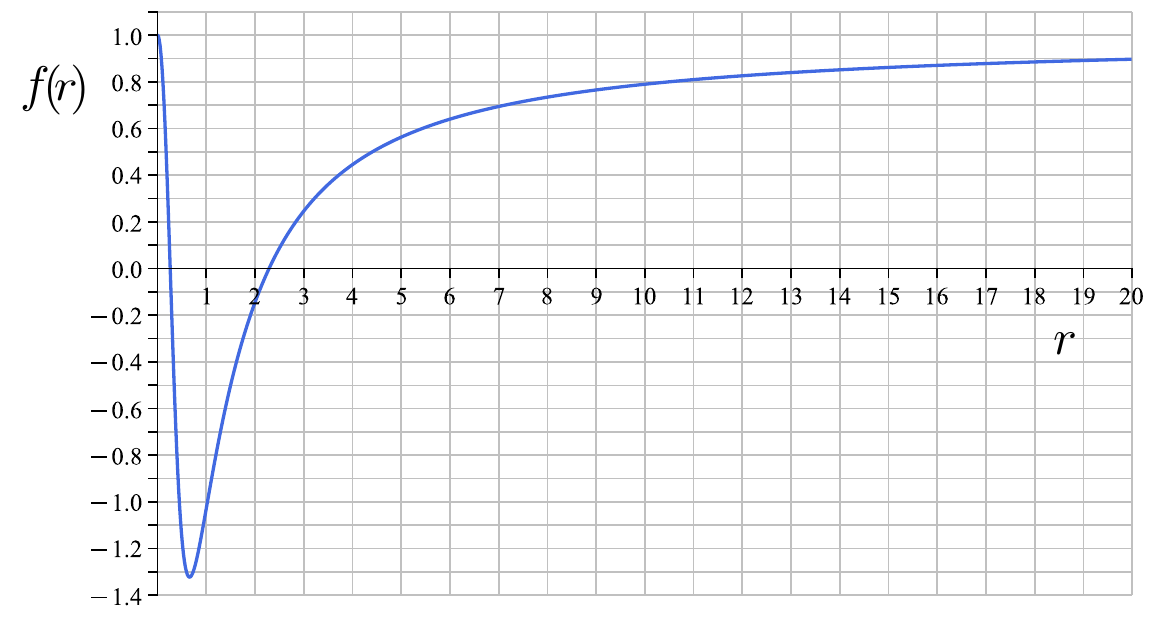}
   \caption{Profiles of the metric function $f(r)$ for various values of the radial coordinate $r$, with fixed mass and magnetic charge parameters $m = q = 1$. The behavior of $f(r)$ governs the horizon structure and the causal properties of the spacetime. The function satisfies $f(0) = 1$, indicating a regular (non-singular) BH solution at the origin. Furthermore, $f(r) \to 1$ as $r \to \infty$, implying that the spacetime is asymptotically flat.}

    \label{metricplot}
\end{figure}

Figure~\ref{figisfull} presents the embedding diagrams illustrating the spatial geometry of the NRCBH for various magnetic charge values with $M = 1$. The visualization displays the unique throat structure, transitioning from the typical Schwarzschild geometry at $q = 0$ (panel a) to more intricate formations with increasing magnetic charge. For small magnetic charges ($q = 0.5$ to $q = 2$, panels b-c), the throat narrows significantly, reflecting the presence of dual horizons with the inner horizon creating a more pronounced constriction. As $q$ approaches the intermediate range ($q = 2.5$ to $q = 3.1$, panels d-g), the geometry of the throat exhibits a gradual trend of widening, consistent with the behavior of the outer horizon observed in Table~\ref{istable}. Near the extremal limit ($q = 3.7$ to $q = 3.79$, panels h-i), the embedding diagrams show a dramatic narrowing of the throat as the inner and outer horizons converge, visually demonstrating the approach to extremality, where the structure of the horizon becomes increasingly constrained before finally disappearing beyond $q_{\text{ext}} \approx 3.79$.

At large radial distances, the expansion of the metric function yields
\begin{equation}
f(r) \sim 1 - \frac{2M}{r} + \frac{q^2}{r^2} + \frac{M^2 q^2 - 8q^4}{4r^4} + \mathcal{O}(r^{-5}), \label{isninf}
\end{equation}
which agrees with the Reissner–Nordström form up to subleading corrections. Near the origin, the geometry approaches a de Sitter core (see also Fig. \ref{metricplot}), with
\begin{equation}
f(r) \sim 1 - \frac{15r^2}{q^2} + \mathcal{O}(r^4), \label{isnsig}
\end{equation}
which guarantees that the curvature invariants remain finite. Figure~\ref{metricplot} illustrates the radial profile of the metric function $f(r)$ for the NRCBH with parameters $M = q = 1$, demonstrating the key features that distinguish this solution from singular BHs. The function exhibits complete regularity at the origin with $f(0) = 1$, confirming the absence of singularities and validating the de Sitter core behavior predicted by the near-origin expansion in Eq.~\eqref{isnsig}. The metric function displays two distinct zeros corresponding to the inner and outer horizons at $r_h \approx 0.26$ and $r_h \approx 2.29$ respectively, consistent with the non-extremal configuration expected for $q = 1$ as shown in Table~\ref{istable}. As $r \to \infty$, the function asymptotically approaches unity, confirming the spacetime's asymptotic flatness and agreement with the Reissner-Nordstr\"{o}m form in the large-distance limit. The smooth, monotonic behavior between the horizons and the absence of pathological features throughout the entire radial domain underscore the physical viability of this regular BH solution.

In particular, the Ricci scalar is computed as follows.
\begin{multline}
R(r) = 12 q^2 \bigg[ \frac{q^2}{(q^2 + r^2)^3} + \frac{24 M^3}{(M q^2 + 8 r^3)^3}\\
- \frac{8 M^2}{(M q^2 + 8 r^3)^2} - \frac{2 q^4}{(q^2 + r^2)^4} \bigg],    
\end{multline}
which reaches its global maximum at the origin:
\begin{equation}
\lim_{r \to 0} R(r) = \frac{180}{q^2}.
\end{equation}
Similarly, the Kretschmann scalar $K = R_{\mu\nu\rho\sigma} R^{\mu\nu\rho\sigma}$ satisfies
\begin{equation}
\lim_{r \to 0} K(r) = \frac{5400}{q^4},
\end{equation}
Signaling complete regularity in the deep interior. These expressions also demonstrate that both $R \to 0$ and $K \to 0$ as $q \to \infty$, in line with asymptotic flatness.

The geometry is sourced by a nonlinear magnetic field, and due to the lack of invertibility between the field strength and its dual, the electromagnetic contribution is described by a Hamiltonian density $\mathcal{H}(\mathcal{P})$, where $\mathcal{P} = \frac{1}{4} P_{\mu\nu}P^{\mu\nu} = -\frac{q^2}{2r^4}$. The explicit form of the Hamiltonian is given by
\begin{multline}
\mathcal{H}(\mathcal{P}) = \frac{3(-2q^2\mathcal{P})^{3/2} - (-2q^2\mathcal{P})}{2q^2\left( \sqrt{-2q^2\mathcal{P}} + 1 \right)^3}\\
-\frac{24(-2q^2\mathcal{P})^{3/2}}{q^2 \left[ (-2q^2\mathcal{P})^{3/4} + 16s \right]^2},
\end{multline}
with $s = q/(2M)$. In the weak-field limit $\mathcal{P} \to 0$, the Hamiltonian reduces to the linear Maxwellian form:
\begin{equation}
\mathcal{H}(\mathcal{P}) \to \mathcal{P}.
\end{equation}
Although the corresponding Lagrangian $\mathcal{L}(\mathcal{F})$ is not accessible through a global Legendre transformation due to the non-invertibility in this model, the magnetic realization remains physically consistent, particularly near the origin where regularity is most sensitive.

Solving the condition $f(r_h) = 0$ yields the structure of the event horizon. For this solution, two horizons exist when the magnetic charge satisfies
\begin{equation}
q < q_{\text{ext}} = 2.537862 M,
\end{equation}
and the geometry becomes extremal at $q = q_{\text{ext}}$. This is in contrast with the Reissner–Nordström case, where extremality occurs at $q = M$, and reflects the influence of the non-linear source on supporting more massive or highly charged configurations without singularity formation.

The energy conditions are evaluated assuming the source behaves as an anisotropic fluid. The energy density is given by
\begin{equation}
\rho = \frac{q^2 r}{4\pi} \left[ \frac{48 M^2}{(M q^2 + 8 r^3)^2} + \frac{r^2 - 3 q^2}{(q^2 + r^2)^3} \right],
\end{equation}
which remains non-negative for $q \leq 1.4528 M$. Furthermore, condition $\rho + p_2 \geq 0$ is fulfilled throughout spacetime if $q \leq 1.2669 M$, ensuring that the Weak Energy Condition (WEC) holds globally for this range. The Dominant Energy Condition (DEC), requiring $\rho - p_2 \geq 0$, is satisfied from the horizon to infinity when $0.3885 M \leq q \leq 1.1672 M$, while the Strong Energy Condition (SEC), involving the combination $\rho + p_1 + p_2 + p_3 \geq 0$, is satisfied outside the horizon provided $q \leq 1.0265 M$. These bounds indicate a well-controlled matter content that obeys classical energy requirements without invoking exotic behavior.

\subsection{Hawking Temperature from Quantum Tunneling Perspective}
To determine the Hawking temperature of the composite BH configuration, we utilized the tunneling approach pioneered by Parikh and Wilczek \cite{parikh2000hawking}, which offers a semi-classical framework to model particle emission as a quantum tunneling process through the event horizon. The method necessitates reformulating the metric in Painlevé-type coordinates to remove coordinate singularities at the horizon:
\begin{equation}
t \rightarrow t - \int \frac{\sqrt{1 - f(r)}}{f(r)}, dr,
\end{equation}
yielding the regularized metric:
\begin{equation}
ds^2 = -f(r) dt^2 + 2 \sqrt{1 - f(r)} dt dr + dr^2 + r^2 d\Omega^2.
\end{equation}

In these coordinates, the outgoing radial null geodesics are governed by the equation:
\begin{equation}
\dot{r} = \frac{dr}{dt} = 1 - \sqrt{1 - f(r)}.
\end{equation}

Close to the event horizon at $r = r_h$, the function $f(r)$ can be linearized via a first-order Taylor expansion.
\begin{equation}
f(r) \approx f'(r_h)(r - r_h),
\end{equation}
which implies the approximate form:
\begin{equation}
\dot{r} \approx 1 - \sqrt{1 - f'(r_h)(r - r_h)} \approx \frac{1}{2} f'(r_h)(r - r_h).
\end{equation}

The tunneling probability is associated with the imaginary part of the action for a particle traversing the horizon, given by:
\begin{equation}
\text{Im}(S) = \text{Im} \int_{r_{\text{in}}}^{r_{\text{out}}} p_rdr = \text{Im} \int_{r_{\text{in}}}^{r_{\text{out}}} \int_0^\omega \frac{d\omega'}{\dot{r}} dr.
\end{equation}

By substituting the near-horizon approximation for $\dot{r}$, we arrive at:
\begin{equation}
\text{Im}(S) \approx \int_0^\omega \int_{r_{\text{in}}}^{r_{\text{out}}} \frac{2}{f'(r_h)(r - r_h)} dr d\omega'.
\end{equation}

This integral features a simple pole at $r = r_h$, and its evaluation via the residue theorem yields:
\begin{equation}
\int_{r_{\text{in}}}^{r_{\text{out}}} \frac{2}{f'(r_h)(r - r_h)} dr = \frac{2\pi i}{f'(r_h)},
\end{equation}

leading to the final form of the imaginary part of the action:
\begin{equation}
\text{Im}(S) = \int_0^\omega \frac{2\pi}{f'(r_h)} d\omega' = \frac{2\pi \omega}{f'(r_h)}.
\end{equation}

Accordingly, the tunneling rate is found as:
\begin{equation}
\Gamma \sim e^{-2\text{Im}(S)} = \exp\left(-\frac{4\pi \omega}{f'(r_h)}\right).
\end{equation}

Comparing this expression with the Boltzmann factor $e^{-\omega/T}$, we extract the Hawking temperature as:

\begin{multline}
T_H = \frac{f'(r_h)}{4\pi} = -\frac{17 r \,M^{2} q^{8}}{2 \left(M \,q^{2}+8 r^{3}\right)^{2} \left(q^{2}+r^{2}\right)^{3} \pi}\\-\frac{47 r^{3} M^{2} q^{6}}{2 \left(M \,q^{2}+8 r^{3}\right)^{2} \left(q^{2}+r^{2}\right)^{3} \pi}-\frac{24 r^{5} M^{2} q^{4}}{\left(M \,q^{2}+8 r^{3}\right)^{2} \left(q^{2}+r^{2}\right)^{3} \pi}\\-\frac{8 r^{7} M^{2} q^{2}}{\left(M \,q^{2}+8 r^{3}\right)^{2} \left(q^{2}+r^{2}\right)^{3} \pi}+\frac{24 r^{4} M \,q^{6}}{\left(M \,q^{2}+8 r^{3}\right)^{2} \left(q^{2}+r^{2}\right)^{3} \pi}\\+\frac{104 r^{6} M \,q^{4}}{\left(M \,q^{2}+8 r^{3}\right)^{2} \left(q^{2}+r^{2}\right)^{3} \pi}+\frac{96 r^{8} M \,q^{2}}{\left(M \,q^{2}+8 r^{3}\right)^{2} \left(q^{2}+r^{2}\right)^{3} \pi}\\+\frac{32 r^{10} M}{\left(M \,q^{2}+8 r^{3}\right)^{2} \left(q^{2}+r^{2}\right)^{3} \pi}-\frac{32 r^{7} q^{4}}{\left(M \,q^{2}+8 r^{3}\right)^{2} \left(q^{2}+r^{2}\right)^{3} \pi}\\+\frac{32 r^{9} q^{2}}{\left(M \,q^{2}+8 r^{3}\right)^{2} \left(q^{2}+r^{2}\right)^{3} \pi}.\label{hawking0}
\end{multline}

\subsection{GUP-Induced Corrections to the Temperature of NRCBH} \label{sec:gup_temp}

GUP emerges as a natural extension of Heisenberg’s uncertainty relation when quantum gravitational effects are taken into account. By introducing a minimal measurable length, GUP modifies the canonical position-momentum commutation relations and has found numerous applications in BH thermodynamics \cite{sucu2025nonlinear}, especially in exploring quantum corrections to the entropy and temperature of charged or rotating configurations. In this section, we investigate how the presence of GUP modifies the Hawking temperature of NRCBH. For this purpose, we work with the spacetime metric introduced earlier in Eq.\eqref{ismetric} and incorporate GUP corrections into the Klein-Gordon equation for a scalar field, following the approach of \cite{sucu2023gup}:
\begin{multline}
-(i\hbar)^2 \partial_t^2 \Phi = \left[(i\hbar)^2 \nabla^i \nabla_i + m_p^2\right] \\
\times \left[1 - 2 \beta_{GUP} \left((i\hbar)^2 \nabla^i \nabla_i + m_p^2\right)\right] \Phi,
\label{eq:gup_kg}
\end{multline}
where $m_p$ is the mass of the test particle and $\beta_{GUP}$ denotes the deformation parameter that quantifies the GUP effect. To extract the quantum tunneling behavior, we apply the WKB approximation and adopt the following ansatz:
\begin{equation}
\Phi(t,r,\theta,\phi) = \exp\left(\frac{i}{\hbar} \mathcal{S}(t,r,\theta,\phi)\right), \label{eq:wkb_ansatz}
\end{equation}
where $\mathcal{S}(t,r,\theta,\psi)$ stands for the classical action. Using the Hamilton-Jacobi approach \cite{parikh2000hawking}, we decompose the action as follows:
\begin{equation}
\mathcal{S}(t,r,\theta,\phi) = -Et + \mathcal{W}(r) + j\phi + \text{const}, \label{eq:hj_action}
\end{equation}
with $E$ and $j$ representing the energy and angular momentum of the tunneling particle, respectively. Near the horizon, we employ the near-horizon expansion of the metric function $f(r)$ as specified in Eq.\eqref{ismetric}.

The quantum tunneling probability, interpreted as the semi-classical emission rate, is found to be:
\begin{equation}
\Gamma \sim \exp\left(-4  \text{Im}  \mathcal{W}(r_h)\right) = \exp\left(\frac{E}{T_{GUP}}\right).
\end{equation}

Solving the modified Klein-Gordon equation under the GUP framework gives the following expression for the imaginary part of the radial action:
\begin{equation}
\mathcal{W}(r_h) = \frac{i \pi E}{f'(r_h) \sqrt{1 - 2 \beta m_p^2}}, \label{eq:w_solution}
\end{equation}
which leads to the corrected BH temperature under GUP considerations:
\begin{equation}
T_{GUP} = T_H \sqrt{1 - 2 \beta m_p^2}. \label{eq
gup_temp}
\end{equation}

As expected, in the limit $\beta \to 0$, the above expression smoothly reduces to the conventional Hawking temperature given earlier in Eq. \eqref{hawking0}, confirming the consistency with the standard semiclassical result.

{\color{black}
\section{Gravitational Lensing and Plasma-Mediated Light Deflection in NRCBH Spacetime} \label{isec3}

The study of light deflection in the gravitational field of NRCBHs provides crucial insights into the observational signatures of these exotic objects and offers a direct probe of the nonlinear electromagnetic effects that distinguish them from classical BH solutions. In this section, we investigate the weak deflection of light rays propagating through the NRCBH spacetime, employing the powerful GBT approach to derive analytical expressions for the deflection angle. Furthermore, we examine how astrophysical plasma environments modify these lensing effects, incorporating frequency-dependent refractive indices that arise from the interaction between photons and the ionized medium surrounding realistic BH systems.

The theoretical framework for analyzing photon trajectories in curved spacetime begins with the recognition that light rays follow null geodesics in the effective optical geometry. For the NRCBH metric given by Eq.~\eqref{ismetric}, we treat the photon path as satisfying the null condition $ds^2 = 0$, which allows us to extract the optical metric from the spatial components of the line element. This procedure yields the effective two-dimensional optical metric $\gamma_{ij}$ that governs photon propagation in the equatorial plane $\theta = \pi/2$:

\begin{equation}
dt^2 = \gamma_{ij} dx^i dx^j 
\end{equation}

To facilitate the application of the GBT, we introduce a tortoise-like coordinate transformation defined by $dr^* = \frac{dr}{f(r)}$, which recasts the metric into a conformally flat form that is more amenable to curvature calculations. This transformation yields:

\begin{equation}
dt^2 = dr^{*2} + \tilde{f}^2(r^*) d\phi^2,
\end{equation}
where the radial function becomes $\tilde{f}(r^*) = \sqrt{\frac{r^2}{f(r)}}$. This conformal representation is particularly valuable because it allows us to compute the Gaussian curvature $\mathcal{K}$ of the optical manifold using standard differential geometric techniques.

For NRCBH geometries arising from NED, the Gaussian curvature of the optical surface is determined by the Ricci scalar of the two-dimensional optical geometry. The calculation proceeds by recognizing that:

\begin{equation}
\mathcal{K} = \frac{R}{2}
\end{equation}
where $R$ represents the Ricci scalar computed on the optical surface. Through careful calculation of the metric components and their derivatives, we obtain the asymptotic expansion of the Gaussian curvature:

\begin{multline}
    \mathcal{K} \approx -\frac{2 M}{r^{3}}+\frac{3 q^{2}}{r^{4}}+\frac{5 M^{2} q^{2}}{2 r^{6}}-\frac{18 q^{4}}{r^{6}}+\frac{3 M^{2}}{r^{4}}\\-\frac{6 M \,q^{2}}{r^{5}}-\frac{9 M^{3} q^{2}}{2 r^{7}}+\frac{36 M \,q^{4}}{r^{7}}+\mathcal{O}(1/r^8)
\end{multline}

This expression reveals the intricate interplay between gravitational ($M$-dependent) and electromagnetic ($q$-dependent) contributions to the spacetime curvature, with the magnetic charge introducing both attractive and repulsive terms that depend on the radial distance from the NRCBH.

The GBT provides a rigorous mathematical framework for computing deflection angles by relating the total curvature of a domain to its topological properties. When applied to a compact domain $D$ bounded by a piecewise smooth curve $\partial D$, the theorem states:

\begin{equation}
\iint_D \mathcal{K} dS + \oint_{\partial D} \kappa \, dt + \sum_i \beta_i = 2\pi \chi(D),
\end{equation}
where $\kappa$ denotes the geodesic curvature along the boundary, $\beta_i$ are exterior angles at corners, and $\chi(D)$ is the Euler characteristic of the domain. For our lensing analysis, we consider a region $\tilde{D}$ bounded by the light trajectory $C_1$ and a large circular arc $C_R$. Since the light ray follows a geodesic with $\kappa(C_1) = 0$ and the domain has Euler characteristic $\chi(\tilde{D}) = 1$, the GBT simplifies to:
\begin{equation}
\iint_{\tilde{D}} \mathcal{K} dS + \int_{C_R} \kappa \, dt = \pi.
\end{equation}

The geodesic curvature on the arc $C_R$ is computed using the Christoffel symbols of the optical metric:
\begin{equation}
\kappa(C_R) = \Gamma^r_{\phi \phi} \left( \frac{d\phi}{dt} \right)^2 = -\tilde{f}(r^*) \frac{d\tilde{f}}{dr^*}.
\end{equation}

In the asymptotic limit $R \to \infty$, the boundary contribution approaches $\kappa(C_R) dt \to d\phi$, which leads to the fundamental relation:

\begin{equation}
\iint_{\tilde{D}_{R\to\infty}} \mathcal{K} dS + \int_0^{\pi + \alpha} d\phi = \pi.
\end{equation}

From this expression, we can isolate the total deflection angle $\alpha$. The area element corresponding to the optical geometry takes the form:
\begin{equation}
dS = \frac{r}{f^{3/2}(r)} \, dr \, d\phi.
\end{equation}

Under the weak field approximation, which is valid for photon trajectories that pass at large impact parameters, we assume a nearly straight path parameterized by $r(\phi) \approx b/\sin\phi$, where $b$ represents the impact parameter. This approximation allows us to express the deflection angle as:
\begin{equation}
\Theta = - \int_0^{\pi} \int_{b/\sin\phi}^{\infty} \mathcal{K} \, dS.
\end{equation}

Performing the integration with the NRCBH curvature expression yields:
\begin{multline}
    \Theta \approx \frac{45 M^{4} q^{2} \pi}{64 b^{6}}+\frac{3 M^{2} \pi}{4 b^{2}}+\frac{93 M^{2} q^{2} \pi}{64 b^{4}}-\frac{45 q^{4} M^{2} \pi}{8 b^{6}}\\-\frac{3 \pi  q^{2}}{4 b^{2}}+\frac{27 q^{4} \pi}{16 b^{4}}-\frac{4 M^{3}}{b^{3}}-\frac{16 M^{3} q^{2}}{25 b^{5}}+\frac{4 M}{b}-\frac{4 M \,q^{2}}{3 b^{3}}+\frac{96 M \,q^{4}}{25 b^{5}}
\end{multline}
\begin{figure}
    \centering
    \includegraphics[width=0.5\textwidth]{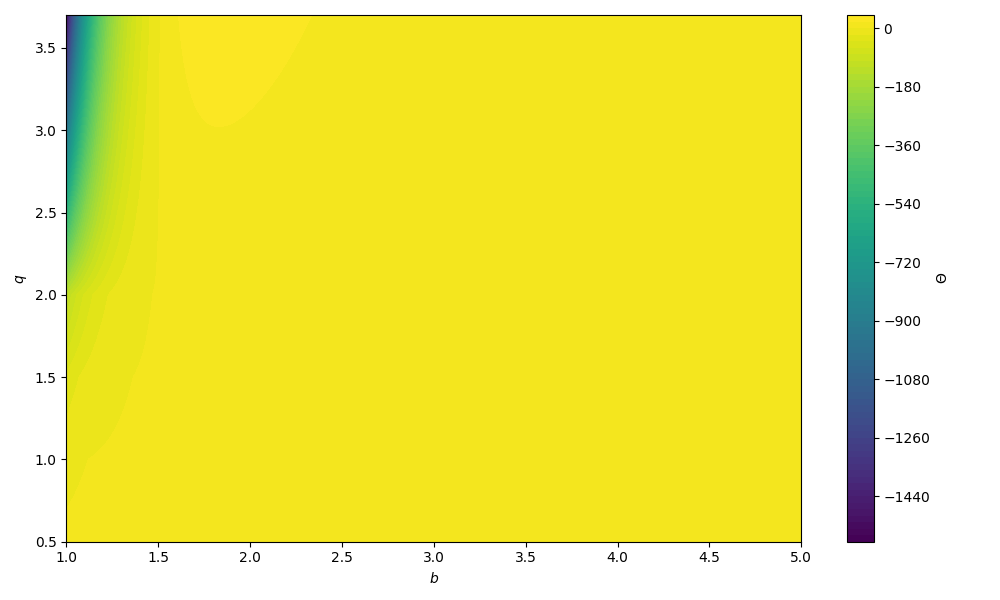}
    \caption{Plot of the deflection angle $\Theta$ as a function of the impact parameter $b$ for various values of the black hole charge $q$. As $b$ increases, $\Theta$ approaches zero, indicating weaker gravitational deflection at larger distances. Higher charge values  significantly alter the bending behavior, with large $q$ leading to negative deflection angles at small $b$, highlighting the influence of electromagnetic repulsion on photon trajectories around charged black holes.
}
    \label{nonplasmaa}
\end{figure}

Figure~\ref{nonplasmaa} illustrates the profound impact of magnetic charge on light deflection in NRCBH spacetimes. The behavior of the deflection angle $\Theta$ as a function of impact parameter $b$ reveals several physically significant features. For the uncharged case ($q=0$), corresponding to the Schwarzschild limit, the deflection remains positive and decreases monotonically with increasing impact parameter, as expected from classical general relativity. However, as the magnetic charge increases, the electromagnetic repulsion begins to compete with gravitational attraction, leading to a dramatic modification of the lensing behavior. For large charge values such as $q=3.7$, the deflection angle becomes negative at small impact parameters, indicating that the electromagnetic repulsion dominates over gravitational attraction in the strong-field regime. This phenomenon represents a unique signature of magnetically charged BHs that could potentially be observed through high-precision astrometric measurements or gravitational microlensing studies.

The incorporation of astrophysical plasma effects introduces additional complexity to the light propagation analysis, as realistic BH environments are typically embedded in ionized media that modify the effective refractive index experienced by photons. We model these effects through a position-dependent refractive index $n(r)$ that depends on the local plasma frequency $\omega_p(r)$ and the photon frequency at infinity $\omega_0$ \cite{sucu2025dynamics}:
\begin{equation}
n(r) = \sqrt{1 - \frac{\omega_p^2(r)}{\omega_0^2} f(r)}.
\end{equation}

This refractive index modifies the optical metric structure, leading to:
\begin{equation}
dt^2 = n^2(r) \left[ \frac{1}{f^2(r)} dr^2 + \frac{r^2}{f(r)} d\phi^2 \right].
\end{equation}

The Gaussian curvature of the plasma-modified optical geometry, denoted $\tilde{\mathcal{K}}$, requires careful calculation of the metric determinant and curvature tensor components:
\begin{equation}
\tilde{\mathcal{K}} = \frac{\mathcal{R}_{r\phi r\phi}}{\det(g^{\text{opt}})} = \frac{\mathcal{R}_{r\phi r\phi}}{n^4(r) \, r^2 / f^3(r)},
\end{equation}

For weak plasma effects, we introduce the dimensionless parameter $\xi = \omega_p^2 / \omega_0^2$ and expand the curvature to first order:
\begin{multline}
\tilde{\mathcal{K}} \approx -\frac{12 M^{3} \xi}{ r^{5}}+\frac{5 M^{2} q^{2}}{2 r^{6}}+\frac{3 M^{2}}{r^{4}}-\frac{18 q^{4}}{r^{6}}\\+\frac{12 M^{2} \xi}{ r^{4}}-\frac{6 M \,q^{2}}{r^{5}}-\frac{3 M \xi}{ r^{3}}-\frac{26 q^{4} \xi}{ r^{6}}+\frac{5 q^{2} \xi}{ r^{4}}\\+\frac{73 M^{2} q^{2} \xi}{2  r^{6}}-\frac{26 M \,q^{2} \xi}{ r^{5}}-\frac{2 M}{r^{3}}+\frac{3 q^{2}}{r^{4}}+\mathcal{O}(1/r^7)
\end{multline}

The plasma-modified area element becomes:
\begin{equation}
dS = \left(r - \frac{\omega_p^2(r)}{\omega_0^2}\right) dr \, d\phi,
\end{equation}
and the deflection angle in the plasma environment is:
\begin{equation}
\tilde{\alpha} = - \int_0^{\pi} \int_{b/\sin\phi}^{\infty} \tilde{\mathcal{K}} \, dS.
\end{equation}

The resulting plasma-corrected deflection angle is:
\begin{multline}
    \tilde{\alpha} \approx \frac{27 M^{4} \pi  \xi}{8  b^{4}}-\frac{3 M^{2} \pi  \xi}{4  b^{2}}+\frac{3 M^{2} \pi}{4 b^{2}}+\frac{249 M^{2} \pi  q^{2} \xi}{64 b^{4}}\\+\frac{93 M^{2} \pi  q^{2}}{64 b^{4}}-\frac{5 \pi  q^{2} \xi}{4  b^{2}}-\frac{3 \pi  q^{2}}{4 b^{2}}+\frac{39 \pi  q^{4} \xi}{16  b^{4}}+\frac{27 \pi  q^{4}}{16 b^{4}}\\-\frac{32 M^{3} \xi}{3  b^{3}}-\frac{4 M^{3}}{b^{3}}-\frac{584 M^{3} q^{2} \xi}{25  b^{5}}-\frac{8 M^{3} q^{2}}{5 b^{5}}+\frac{6 M \xi}{ b}\\+\frac{4 M}{b}+\frac{44 q^{2} \xi M}{9  b^{3}}-\frac{4 M \,q^{2}}{3 b^{3}}+\frac{416 q^{4} \xi M}{25  b^{5}}+\frac{288 q^{4} M}{25 b^{5}}+\mathcal{O}(1/b^6)
\end{multline}
\begin{figure}
    \centering
    \includegraphics[width=0.5\textwidth]{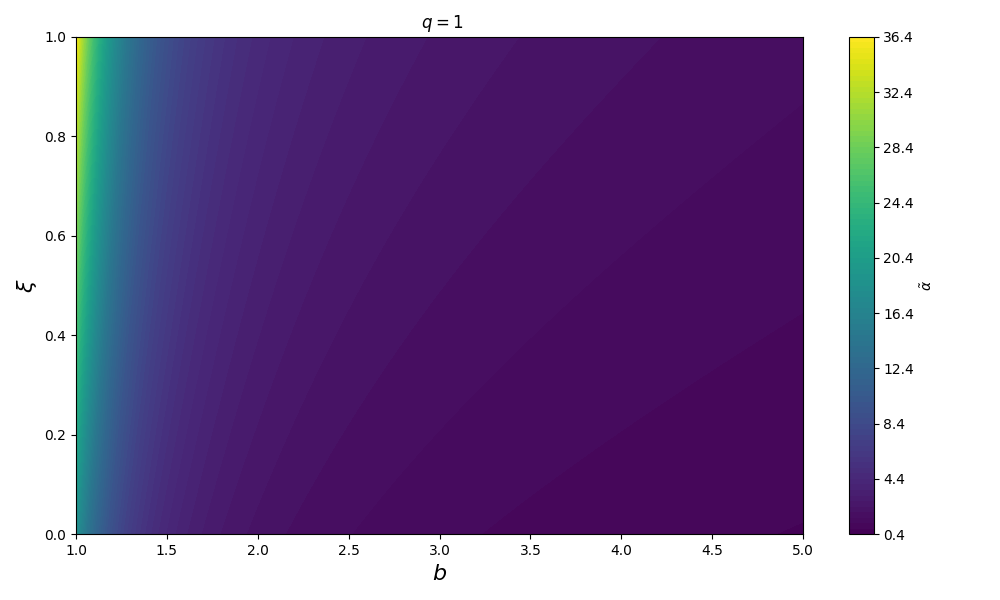}
    \caption{Density plot of the plasma deflection angle $\tilde{\alpha}$ as a function of the impact parameter $b$ and the Lorentz-violating parameter $\xi$ for fixed charge $q=1$.}
    \label{plasmaa1}
\end{figure}

Figure~\ref{plasmaa1} provides a comprehensive visualization of how plasma effects modify the deflection angle for a fixed magnetic charge $q=1$. The density plot reveals the interaction between the impact parameter $b$ and the plasma parameter $\xi$, showing how increasing the plasma density (larger $\xi$) generally enhances the deflection effect. This plasma-induced modification represents an important astrophysical consideration, as most realistic BH environments contain significant amounts of ionized matter that can substantially alter the observed lensing signatures. The frequency dependence inherent in the plasma effects also suggests that multi-wavelength observations could provide additional constraints on both the BH parameters and the properties of the surrounding plasma medium.

\section{Orbital Dynamics and Quasi-Periodic Oscillations in NRCBH Spacetimes} \label{isec4}

The study of orbital dynamics around NRCBHs provides fundamental insights into the characteristic frequencies that govern the motion of test particles in these exotic gravitational environments. These frequencies are not merely theoretical constructs but represent directly observable quantities that manifest as QPOs in the X-ray emissions from accreting matter around astrophysical BHs. The unique electromagnetic structure arising from NED in these regular spacetimes introduces distinctive modifications to the orbital properties that can potentially serve as observational signatures to distinguish NRCBHs from their classical counterparts.

The analysis of circular orbital motion begins with the recognition that neutral test particles follow timelike geodesics in the NRCBH spacetime defined by the metric function given in Eq.~\eqref{ismetric}. For stable circular orbits in the equatorial plane, the fundamental quantity of interest is the angular velocity $\Omega$, which characterizes the rotation rate of a particle as measured by observers at spatial infinity. This angular frequency is intimately connected to the ratio of coordinate changes and represents the most basic dynamical property of circular motion:
\begin{equation}
\Omega = \frac{d\phi}{dt}.
\end{equation}

For spherically symmetric and static spacetimes, the derivation of the angular velocity for circular geodesics can be accomplished through the standard variational approach applied to the effective potential. The resulting expression involves derivatives of the metric components and takes the elegant form:
\begin{equation}
\Omega = \sqrt{ \frac{-\partial_r g_{tt}}{\partial_r g_{\phi\phi}} },
\end{equation}
where $g_{tt}$ and $g_{\phi\phi}$ represent the temporal and azimuthal components of the metric tensor, respectively. For the NRCBH spacetime under consideration, we have $g_{tt} = -f(r)$ and $g_{\phi\phi} = r^2$, which substantially simplifies the angular velocity expression to:
\begin{equation}
\Omega = \sqrt{ \frac{f'(r)}{2 r} },
\end{equation}
where the prime notation denotes differentiation with respect to the radial coordinate $r$. This expression reveals the direct dependence of the orbital frequency on the local curvature properties of the spacetime, as encoded in the derivative of the metric function.

Inserting the explicit form of the NRCBH metric function from Eq.~\eqref{ismetric} results in a more intricate yet physically insightful expression for the angular velocity. After performing the necessary algebraic manipulations, we obtain:
\begin{equation}
\Omega = \sqrt{\frac{A}{B}},
\end{equation}
where the numerator and denominator are given by the following expressions:
\begin{multline}
    A=-17 M^{2} q^{8}-47 M^{2} q^{6} r^{2}-48 M^{2} q^{4} r^{4}-16 M^{2} q^{2} r^{6}\\+48 M \,q^{6} r^{3}+208 M \,q^{4} r^{5}+192 M \,q^{2} r^{7}\\+64 M \,r^{9}-64 q^{4} r^{6}+64 q^{2} r^{8}
\end{multline}

and
\begin{equation}
    B=\left(M \,q^{2}+8 r^{3}\right)^{2} \left(q^{2}+r^{2}\right)^{3}
\end{equation}

These expressions reveal the intricate interplay between the gravitational mass $M$, the magnetic charge $q$, and the radial coordinate $r$ in determining the orbital characteristics. The mixture of positive and negative terms within the numerator $A$ implies complicated orbital architectures, potentially leading to extremal points in the angular velocity profile that denote unique orbital configurations.

To connect these theoretical results with observational quantities, we express the angular frequency in physically meaningful units. The conversion to Hertz, which represents the natural frequency scale for astrophysical observations, is accomplished through:
\begin{equation}
\nu_\phi = \frac{c^3}{2\pi G M}  \Omega_\phi,
\end{equation}
where $c$ denotes the speed of light and $G$ represents the gravitational constant. This transformation allows for direct comparison with observational data from X-ray binaries and other accreting BH systems, where QPO frequencies are routinely measured with high precision.

The physical behavior of the angular velocity profile exhibits several remarkable features that distinguish NRCBH spacetimes from their classical counterparts. As demonstrated in Figure~\ref{omega}, the charge-dependent modifications to the orbital dynamics manifest in dramatic ways that have profound implications for observational astronomy.

\begin{figure}
    \centering
    \includegraphics[width=0.5\textwidth]{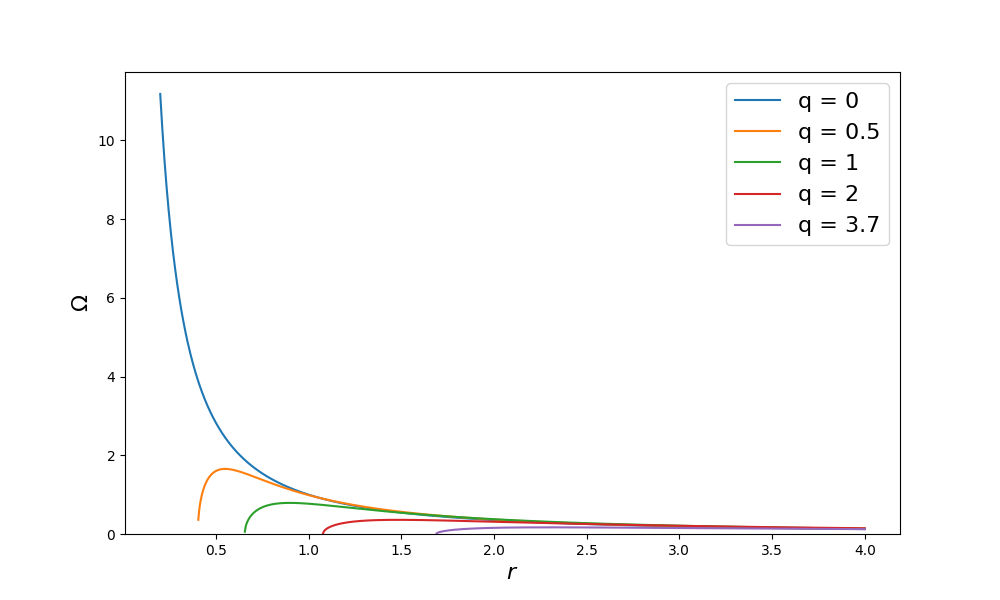}
    \caption{ The Keplerian angular velocity $\Omega(r)$ for different electric charge values $q$ in the NRCBH spacetime. The presence of electric charge modifies the angular velocity profile significantly, introducing a local maximum that shifts outward and diminishes in magnitude as $q$ increases. This behavior reflects the interplay between gravitational attraction and electrostatic repulsion, and has direct implications for the location and stability of circular orbits, as well as for modeling QPOs in charged black hole environments.
}
    \label{omega}
\end{figure}

The analysis of Figure~\ref{omega} reveals several physically significant trends that emerge from the charge-dependent orbital dynamics. For the uncharged case ($q=0$), the angular velocity exhibits the familiar monotonic decrease characteristic of Schwarzschild spacetime, reflecting the weakening of gravitational influence with increasing orbital radius. This behavior aligns with classical expectations from Newtonian dynamics and represents the baseline against which charge-induced modifications can be assessed.

However, the introduction of magnetic charge fundamentally alters the orbital landscape in ways that reflect the underlying physics of NED. For moderate charge values ($q = 0.5, 1, 2$), the angular velocity profile develops a distinctive local maximum that represents a transition between inner and outer orbital regimes. This maximum point is particularly significant from an astrophysical perspective, as it corresponds to orbits where the gravitational and electromagnetic forces achieve a delicate balance that maximizes the orbital frequency. As the magnetic charge increases further ($q = 3.7$), the position of this maximum shifts to larger radial coordinates while its magnitude decreases, indicating that the electromagnetic repulsion increasingly dominates the orbital dynamics at smaller radii. This behavior has profound implications for the interpretation of QPO observations, as the characteristic frequencies associated with these special orbits will be systematically modified in magnetically charged systems compared to their classical counterparts. The existence of these frequency maxima is particularly relevant for understanding high-frequency QPOs observed in accreting BH systems. These phenomena are often attributed to orbital motion near the innermost stable circular orbit (ISCO), and the charge-dependent modifications revealed in our analysis suggest that NRCBH systems could exhibit distinctive QPO signatures that might be distinguishable from those produced by classical BHs.

\section{Thermodynamic Phase Transitions and Joule-Thomson Expansion in NRCBH Spacetimes} \label{isec5}

The investigation of thermodynamic phase transitions in BH physics has emerged as a powerful framework for understanding the stability and critical behavior of gravitational systems. In this context, the JTE provides a particularly insightful probe of isenthalpic processes that reveal fundamental aspects of BH thermodynamics within the extended phase space formalism. For NRCBHs arising from NED, the interplay between gravitational and electromagnetic contributions to the thermodynamic potentials introduces rich phase behavior that can be systematically analyzed through the JTE framework.

The JTE represents a thermodynamic transformation characterized by constant enthalpy, during which the temperature exhibits a characteristic dependence on pressure variations. When adapted to BH physics within the extended phase space paradigm, where the cosmological constant is interpreted as thermodynamic pressure and its conjugate quantity as thermodynamic volume, this process provides a natural mechanism for investigating isenthalpic trajectories in the space of BH parameters. The fundamental quantity governing this behavior is the Joule-Thomson coefficient, which serves as a diagnostic tool for identifying cooling and heating regimes during expansion or compression processes.

The thermodynamic foundation of our analysis begins with the specific heat capacity associated with the NRCBH, which can be derived from the first law of thermodynamics and the relationship between entropy and temperature. The heat capacity is formally defined as:
\begin{equation}
C = T_H \left( \frac{\partial S}{\partial T_H} \right), \label{s25}
\end{equation}
where $S = \pi r_h^2$ represents the Hawking-Bekenstein entropy expressed in terms of the event horizon radius. Through careful calculation of the temperature derivatives and incorporation of the NRCBH metric properties, we obtain the heat capacity:
\begin{equation}
    C\approx -\frac{2 \pi \left(M^{2} q^{2}-2 M \,r^{3}-8 q^{4}+2 q^{2} r^{2}\right) r^{2}}{5 M^{2} q^{2}-4 M \,r^{3}-40 q^{4}+6 q^{2} r^{2}}
\end{equation}

This expression highlights the intricate relationship of heat capacity with mass $M$ and magnetic charge $q$, noting potential sign changes that suggest transitions between thermodynamic stability and instability.

The Joule-Thomson coefficient, which quantifies the temperature response under isenthalpic conditions, is formally defined as:
\begin{equation}
\mu_J = \left( \frac{\partial T_H}{\partial P_C} \right)_H,
\end{equation}
where $P_C$ represents the critical pressure and the subscript $H$ indicates constant enthalpy. This coefficient provides the fundamental criterion for distinguishing between cooling ($\mu_J > 0$) and heating ($\mu_J < 0$) processes during isenthalpic expansion. The mathematical relationship between the Joule-Thomson coefficient and other thermodynamic quantities can be expressed through the alternative formulation:

\begin{equation}
\mu_J = \frac{1}{C} \left[ T_H \left( \frac{\partial V}{\partial T_H} \right)_P - V \right],
\end{equation}
where $C$ denotes the heat capacity at constant pressure and $V$ represents the thermodynamic volume. This expression establishes the connection between the expansion coefficient, thermal properties, and the fundamental geometric characteristics of the BH.

For the NRCBH system under investigation, the explicit calculation of the Joule-Thomson coefficient yields:
\begin{equation}
\mu_J \approx \frac{4 \left(4 M^{2} q^{2}-5 M \,r^{3}-32 q^{4}+6 q^{2} r^{2}\right)   r}{3 \left(-8 q^{4}+\left(M^{2}+2 r^{2}\right) q^{2}-2 M \,r^{3}\right) },
\end{equation}

This result demonstrates the intricate dependence of the expansion behavior on the NRCBH parameters, with both the mass and magnetic charge contributing to the determination of cooling versus heating regimes.

\begin{figure}
    \centering
    \includegraphics[width=0.5\textwidth]{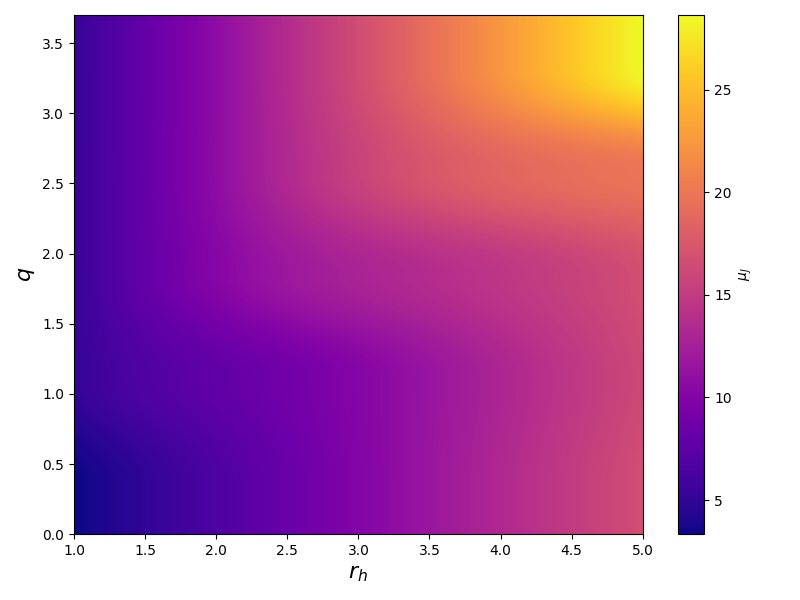}
    \caption{Density plot of the Joule-Thomson coefficient $\mu_J$ as a function of the event horizon radius $r_h$ and the magnetic charge $q$ for NRCBH spacetimes. The color scale represents the magnitude and sign of $\mu_J$, with positive values indicating cooling regions and negative values corresponding to heating regimes during isenthalpic expansion.}
    \label{JTEgraph}
\end{figure}

The comprehensive analysis presented in Figure~\ref{JTEgraph} provides crucial insights into the thermodynamic phase structure of NRCBH spacetimes. The density plot reveals how the Joule-Thomson coefficient $\mu_J$ varies across the parameter space defined by the event horizon radius $r_h$ and magnetic charge $q$. The vertical axis represents the charge parameter, while the horizontal axis corresponds to the horizon radius, with the color scale encoding the magnitude and sign of the expansion coefficient.

The physical interpretation of these results reveals several remarkable features of NRCBH thermodynamics. Regions where $\mu_J > 0$ correspond to cooling behavior during isenthalpic expansion, indicating that the BH temperature decreases as the system undergoes expansion at constant enthalpy. Conversely, regions with $\mu_J < 0$ represent heating regimes where expansion leads to temperature increase, suggesting thermodynamically unstable behavior that could drive phase transitions. The systematic variation of $\mu_J$ across the parameter space demonstrates that NRCBHs with larger charges and event horizons generally exhibit enhanced cooling behavior during JTE processes. This trend suggests that highly charged, massive NRCBHs possess greater thermodynamic stability in the sense that they naturally evolve toward lower-temperature states during expansion. The gradient variations observed in different regions of the parameter space are particularly significant, as they indicate the sensitivity of the thermodynamic response to small parameter changes and may signal the proximity of critical points or phase transition boundaries. From an astrophysical perspective, these JTE properties have important implications for understanding the long-term evolution and stability of magnetically charged BH systems. The cooling behavior exhibited by large, highly charged NRCBHs suggests that such objects would naturally evolve toward more stable configurations through thermodynamic processes. Furthermore, the detailed structure of the $\mu_J$ distribution could provide observational diagnostics for constraining the charge and mass parameters of astrophysical BHs through indirect thermodynamic signatures.

\section{Concluding Remarks} \label{isec6}

In this comprehensive investigation, we conducted a detailed analysis of NRCBHs arising from nonlinear electrodynamics (NED), focusing on their thermodynamic properties, gravitational lensing characteristics, orbital dynamics, and JTE behavior. Our study provided new insights into the observational signatures and physical properties of these exotic gravitational objects, revealing distinctive features that could potentially distinguish them from classical BH solutions in future astrophysical observations.

We began our analysis by examining the geometric and thermodynamic foundation of NRCBHs, establishing the metric function given in Eq.~\eqref{ismetric} that ensures complete regularity throughout the spacetime while maintaining asymptotic flatness. Our investigation of the horizon structure, as detailed in Table~\ref{istable}, revealed that the extremal magnetic charge limit reaches $q_{\text{ext}} \approx 3.79M$, which significantly exceeds the classical Reissner-Nordstr\"{o}m value of $q = M$. This enhancement demonstrates the stabilizing influence of nonlinear electromagnetic effects on the horizon structure and establishes the physical parameter range for viable NRCBH solutions. The embedding diagrams presented in Figure~\ref{figisfull} provided visual confirmation of the smooth geometric transition from the Schwarzschild limit to highly charged configurations, illustrating the characteristic throat evolution that accompanies increasing magnetic charge.

Our thermodynamic analysis employed the quantum tunneling framework pioneered by Parikh and Wilczek to derive the Hawking temperature of NRCBHs. Equation~\eqref{hawking0} highlights the intricate interaction of gravitational and electromagnetic factors in thermal emission. We further incorporated corrections arising from the GUP, obtaining the GUP modified temperature expression seen in Eq.~\eqref{eq
gup_temp}. This result demonstrated how quantum gravitational effects systematically modify the standard Hawking radiation, providing a framework for testing GUP phenomenology through BH thermodynamics. The smooth reduction to the classical limit as $\beta \to 0$ confirmed the consistency of our approach and established the validity of the quantum corrections.

The investigation of gravitational lensing in NRCBH spacetimes constituted a major component of our analysis, where we applied the GBT to derive exact expressions for light deflection angles. Our calculations revealed remarkable charge-dependent behavior, as illustrated in Figure~\ref{nonplasmaa}, where we observed the transition from positive deflection angles in the classical limit to negative values for highly charged configurations. This phenomenon, arising from the competition between gravitational attraction and electromagnetic repulsion, represents a unique observational signature of magnetically charged BHs that could potentially be detected through high-precision astrometric measurements or gravitational microlensing studies. We extended our lensing analysis to incorporate astrophysical plasma effects, recognizing that realistic BH environments typically contain significant amounts of ionized matter. The plasma-modified deflection angle, expressed through the comprehensive formula involving the dimensionless parameter $\xi = \omega_p^2/\omega_0^2$, demonstrated how frequency-dependent refractive indices systematically alter the light propagation characteristics. The density plot presented in Figure~\ref{plasmaa1} revealed the parameter space dependence of these effects, showing how plasma density variations could provide additional observational diagnostics for both BH parameters and the properties of the surrounding medium. These results have important implications for interpreting multi-wavelength observations of accreting BH systems, where plasma effects could significantly modify the apparent lensing signatures.

Our analysis of orbital dynamics and QPOs provided significant insights into the characteristic frequencies governing test particle motion around NRCBHs. The angular velocity expression $\Omega = \sqrt{f'(r)/(2r)}$ and its explicit form involving the expressions for the numerator and denominator terms revealed the intricate dependence on both mass and magnetic charge parameters. Figure~\ref{omega} demonstrated the dramatic charge-induced modifications to the orbital frequency profiles, showing the emergence of characteristic maxima that shift to larger radii and diminish in magnitude as the magnetic charge increases. These frequency maxima correspond to special orbital configurations where gravitational and electromagnetic forces achieve optimal balance, and their locations provide potential diagnostic tools for inferring the magnetic charge of astrophysical BHs through detailed QPO spectroscopy.

Our investigation of JTE properties revealed the remarkable thermodynamic phase structure of NRCBH spacetimes within the extended phase space formalism. The heat capacity calculation and the derivation of the Joule-Thomson coefficient $\mu_J$ provided comprehensive characterization of isenthalpic processes in these systems. The density plot in Figure~\ref{JTEgraph} illustrated the systematic variation of the expansion coefficient across the parameter space defined by horizon radius and magnetic charge, revealing distinct cooling and heating regimes that reflect the underlying thermodynamic stability properties. The physical interpretation of our JTE results demonstrated that NRCBHs with larger charges and event horizons generally exhibit enhanced cooling behavior during expansion processes, suggesting greater thermodynamic stability compared to their classical counterparts. Throughout our investigation, we maintained careful attention to the energy conditions and physical viability of the NRCBH solutions. Our analysis confirmed that the WEC holds globally for magnetic charges $q \leq 1.2669M$, while the DEC is satisfied in the range $0.3885M \leq q \leq 1.1672M$. These constraints establish the physically reasonable parameter space for NRCBH models and provide guidance for observational searches for such exotic objects. The synthesis of our results across multiple physical phenomena - thermodynamics, lensing, orbital dynamics, and phase transitions - revealed a consistent picture of NRCBH observational signatures that could guide current and future observational efforts. 

Looking toward future research directions, our study opens several promising avenues for further investigation \cite{ashtekar2005black,barrau2014planck,tsukamoto2017deflection}. The extension of our lensing analysis to strong-field regimes could reveal additional distinctive signatures of NRCBHs, particularly in the context of photon ring imaging with next-generation telescopes. The incorporation of rotation into the NRCBH framework would provide more realistic models for comparison with astrophysical observations, while the investigation of gravitational wave signatures from extreme mass ratio inspirals could offer new probes of the spacetime geometry. Additionally, the development of more sophisticated plasma models that account for magnetic field effects and non-uniform density distributions could enhance the accuracy of our lensing predictions. Finally, the exploration of quantum field theory effects in NRCBH backgrounds might provide deeper insights into the microscopic origin of Hawking radiation and the information paradox resolution in regular spacetimes.

\acknowledgments 
 E. A. grateful to Professor Paul J. Steinhardt and Princeton University for warm hospitality. E. A. also acknowledges T\"{U}B\.{I}TAK since this study is partially supported by T\"{U}B\.{I}TAK under 2219 Project: "\textit{Studies in the framework of the Chaotic Cyclic Cosmology}".  E. S. and \.{I}. S. acknowledge the academic support of EMU, T\"{U}B\.{I}TAK, SCOAP3, and ANKOS. \.{I}.~S. also thanks COST Actions, specifically projects CA22113, CA21106, and CA23130, for their networking support.

\bibliographystyle{apsrev4-1}
\bibliography{ref}
\end{document}